\documentclass[aps,amssymb,amsmath,prl,twocolumn,showpacs]{revtex4-1}
\usepackage{graphicx}
\usepackage{dcolumn}
\usepackage{bm}
\usepackage{subfigure}

\usepackage{amssymb,amsmath,amsfonts,latexsym,graphicx,verbatim,dsfont}
\usepackage[matrix,frame,arrow]{xy}
\usepackage{epsf}
\usepackage[usenames,dvipsnames]{color} 

\definecolor{LinkColor}{rgb}{0,0,.5}
\usepackage[colorlinks=true,linkcolor=BrickRed,citecolor=LinkColor,urlcolor=LinkColor]{hyperref}
\usepackage{ulem}
\renewcommand{\emph}{\textit}

\definecolor{Myblue}{rgb}{0,0,1}

\newcommand{\B}[1]{\mathbf{#1}}

\newcommand{\I}[1]{\textit{#1}}
\newcommand{\R}[1]{\textrm{#1}}
\newcommand{\ham}{{\mathcal{H}}}

\newcommand{\xd}{\delta}

\newcommand{\qt}{\tau}
\newcommand{\xo}{\omega}
\newcommand{\app}{\approx}
\newcommand{\beq}{\begin{equation}}
\newcommand{\eeq}{\end{equation}}
                  
\newcommand{\bea}{\begin{eqnarray}}
\newcommand{\eea}{\end{eqnarray}}

\newcommand{\non}{\nonumber}
\newcommand{\lb}{\left(}
\newcommand{\rb}{\right)}

\newcommand{\zl}[1]{\label{eqn:#1}}

\newcommand{\zfl}[1]{\label{fig:#1}}
\newcommand{\zfr}[1]{Fig.~\ref{fig:#1}}
\newcommand{\ztl}[1]{\label{table:#1}}
\newcommand{\ztr}[1]{Table~\ref{table:#1}}

\newcommand{\zt}{\times}
\newcommand{\ov}[1]{\overline{#1}}
\newcommand{\tr}[1]{\textrm{Tr}\left\{{#1}\right\}}
\graphicspath{{../}}
\begin{document}
\title{Quantum simulation via filtered Hamiltonian engineering: application to perfect
quantum transport in spin networks}
\author{Ashok Ajoy}
\affiliation{Department of Nuclear Science and Engineering and Research
Laboratory of Electronics,
Massachusetts Institute of Technology, Cambridge, MA, USA}
\author{Paola Cappellaro}
\affiliation{Department of Nuclear Science and Engineering and Research
Laboratory of Electronics,
Massachusetts Institute of Technology, Cambridge, MA, USA}

\begin{abstract}
We propose a method for Hamiltonian engineering in quantum information
processing architectures that requires no local control, but only
relies on collective qubit rotations and field gradients.  The
technique achieves a spatial modulation of the coupling strengths via
a dynamical construction of a weighting function combined with a Bragg
grating.  As an example, we demonstrate how to generate the ideal
Hamiltonian for perfect quantum information transport between two
separated nodes of a large spin network.  We engineer a spin chain
with optimal couplings from a large spin network, such as naturally
occurring in crystals, while decoupling all unwanted interactions. For realistic experimental parameters, our method can be used to drive perfect quantum information transport at
room-temperature.  The Hamiltonian engineering method can be made more robust under coherence and coupling disorder by a novel  apodization scheme. Thus the method is quite general and can be used engineer the Hamiltonian of many complex spin lattices with different topologies and interactions.
\end{abstract}
\maketitle

Controlling the evolution of complex quantum systems has emerged as an
important area of research for its promising practical
applications. The control task can be often reduced to the goal of
Hamiltonian engineering~\cite{Schirmer07} (and also extended to reservoir
engineering~\cite{Verstraete09,Kraus08}).  Hamiltonian
engineering has been used to achieve a variety of tasks, including
quantum computation~\cite{DiVincenzo00b}, improved quantum
metrology~\cite{Cappellaro09b} and protection against decoherence by
dynamical decoupling filtering~\cite{Ajoy11,Biercuk11}. The most
important application is quantum simulation~\cite{Cirac12}, as first
proposed by Feynman~\cite{Feynman82}. The ultimate goal is to achieve
a programmable universal quantum simulator, able to mimic the dynamics
of any system. A possible strategy is to use a quantum computer and
decompose the desired evolution into unitary
gates~\cite{Ajoy12b,Blatt12}. Alternatively, one can use Hamiltonian
engineering by a Suzuki-Trotter factorization of the desired
interaction into experimentally achievable
Hamiltonians~\cite{Lloyd96,Suzuki90}.  However, experimental
implementations of these simulation methods often require \I{local}
quantum control, which is difficult to achieve in large quantum
systems.

Instead, here we present a scheme for Hamiltonian
  engineering that employs only collective
rotations of the qubits and field gradients -- technology readily available
e.g. in magnetic resonance imaging, ion traps~\cite{Blatt12} and optical
lattices~\cite{Simon11}.  We consider a qubit network (eg. \zfr{network}, naturally occurring in a crystal lattice) with an internal
Hamiltonian $H_{\R{int}}$, for example due to the dipolar coupling
between the spins.  The target Hamiltonian $H_{\R{tar}}$ is
engineered from $H_{\R{int}}$ by first ``removing'' unwanted couplings and
then ``modulating'' the remaining coupling strengths to match $H_{\R{tar}}$.  The first step is equivalent to the construction of a
time-domain Bragg grating ${\cal G}_{ij}$, which acts like a sharp
filter, retaining only specific couplings.  Then, the weighting
function $F_{ij}$ allows fine-tuning of the couplings, without the
need for local control.

Hamiltonian engineering has a long history in nuclear magnetic
resonance, as described by coherent averaging~\cite{Haeberlen68} and
field gradients have been proposed to achieve NMR ``diffraction'' in
solid~\cite{Mansfield73a,Mansfield75}.  Although these techniques have
been recently extended in the context of dynamical
decoupling~\cite{Ajoy11} and selective
excitation~\cite{Paravastu06,Alvarez10,Smith12}, our method allows
increased control on the filtering action, thanks to the engineered
grating~\cite{Kashyap99b} obtained with magnetic-field
gradients. Achieving tunability of the filtered coupling strengths
opens the possibility of quantum simulation with only global control.

As an example, we show how to apply this filtered engineering method
to the specific problem of generating an optimal Hamiltonian for
quantum information transfer (QIT).  Linear arrays of spins have been
proposed as quantum \I{wires} to link separated nodes of a spin
network~\cite{Bose03}; engineering the coupling between the spins can
achieve perfect QIT~\cite{Christandl04}.  Finally, we will analyze
experimental requirements to implement the method in existing physical
architectures.

\begin{figure}[t]
 \centering
\includegraphics[width=0.3\textwidth]{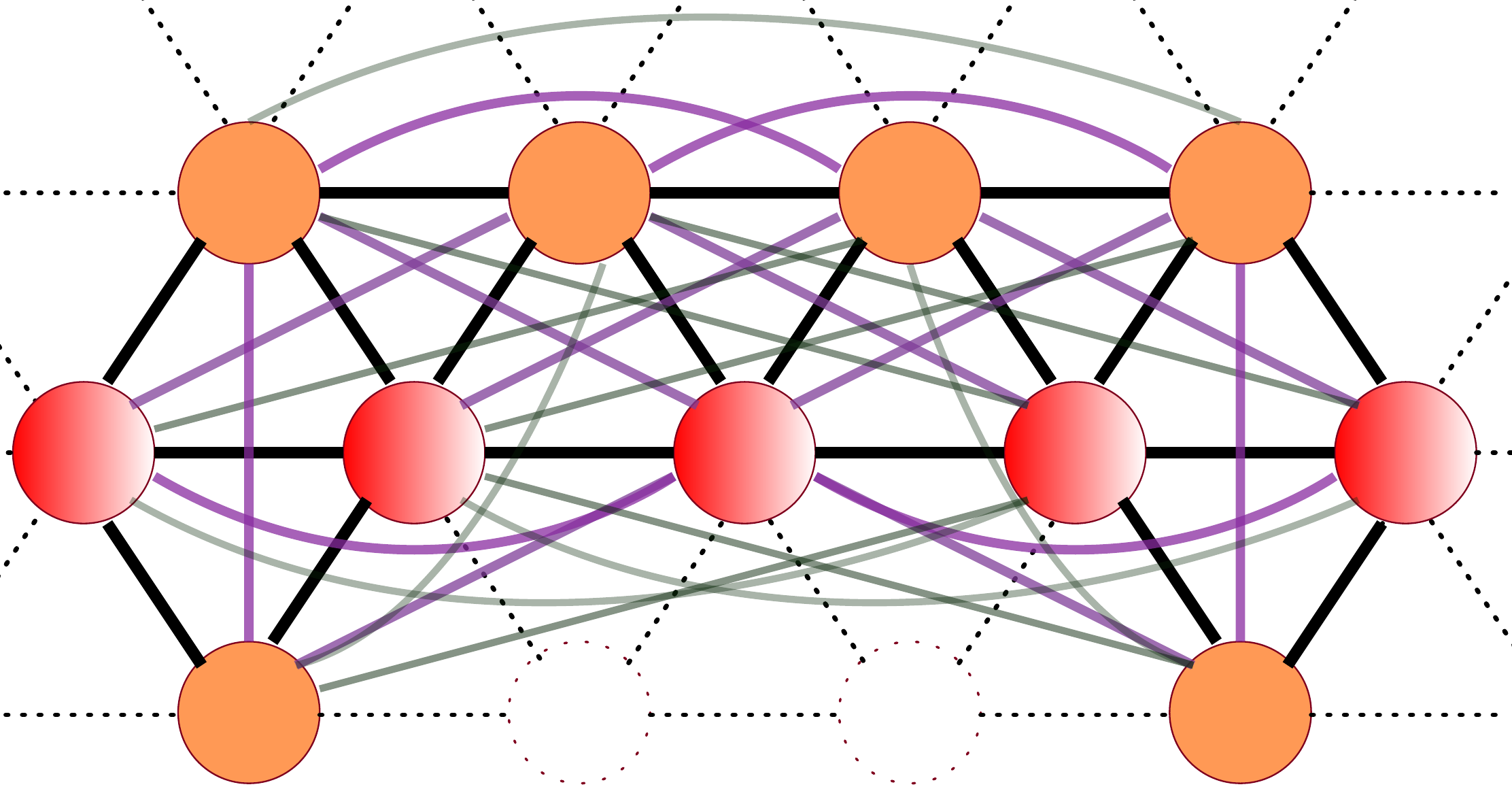}
\caption{A complex spin network in a trigonal planar lattice consisting of a linear chain (red) and
surrounding network (orange). The edges denote
couplings between the spins; only the spins considered in later simulations are depicted. 
A linear magnetic field gradient is aligned in the direction of the chain, such
that the $j^{\R{th}}$ spin frequency is $\xo_j=j\xo$. The filter eliminates NNN couplings as well as off-chain couplings.}
\zfl{network}
\end{figure}
\paragraph{Hamiltonian engineering --}The goals of filtered
Hamiltonian engineering can be summarized as (i) the cancellation of
unwanted couplings -- often next-nearest neighbor (NNN) and other long
range couplings -- and (ii) the engineering of the remaining couplings
to match the desired coupling strengths in $H_{\R{tar}}$ .  We achieve these goals by
dynamically generating tunable and independent grating (${\cal
  G}_{ij}$) and weighting (${F}_{ij}$) functions by means of
collective rotations under a gradient.  The first step is to create the
Hamiltonian operator one wishes to simulate using sequences of
collective pulses.  Although the initial Hamiltonian $H_{\R{int}}$
restricts which operators can be obtained~\cite{SOM}, various control
sequences have been proposed to realize a broad set of
Hamiltoninas~\cite{Baum85,Suter87}.  These multiple pulse
sequences cannot however modulate specific coupling {strengths}, which is
instead our goal -- this can be achieved by evolution under a magnetic field gradient. Consider for example the
isotropic XY Hamiltonian,
$H_{XY}=\sum_{ij}b_{ij}(S_i^xS_j^x+S_i^yS_j^y)$.  {Evolution under the
  propagator} $U(t,\qt)=e^{-iH_z\tau}e^{-iH_{XY}t}e^{iH_z\tau}$, where
$H_z=\sum_i\omega_iS_i^z$ is obtained by a gradient, is equivalent to
evolution under the Hamiltonian 
\beq
  H'_{XY}=\sum_{ij}b_{ij}[(S_i^xS_j^x+S_i^yS_j^y)\cos(\delta\omega_{ij}\tau)\\\qquad\qquad+(S_i^xS_j^y-S_i^yS_j^x)\sin(\delta\omega_{ij}\tau)],
\eeq
where $\delta\omega_{ij}\!=\!\omega_j\!-\!\omega_i$.  The
  modulation can be repeated to obtain a propagator $U_0=\prod_h
  U(t_h,\tau_h)\approx e^{i\ov H T}$ over the total time $T$, where
  the effective Hamiltonian $\ov{H}$ can be calculated e.g. from a
  first order average Hamiltonian theory
  expansion~\cite{Haeberlen68,Haeberlen76}. Given a desired target
  Hamiltonian
  $H_d=\sum_{ij}d^1_{ij}(S_i^xS_j^x+S_i^yS_j^y)+d^2_{ij}(S_i^xS_j^y-S_i^yS_j^x)$,
  we obtain a set of equations in the unknowns $\{t_h,\tau_h\}$ by
  imposing {$\ov{H}= H_d$}.

To simplify the search for the correct timings, we can further apply a
filter that cancels all unwanted couplings and use the equations above to
only determine parameters to engineer the remaining, non-zero, couplings. The filter is
obtained by a dynamical implementation of a Bragg
{grating}: instead of the propagator $U_0$, we evolve
under $N$ cycles (while reducing the times to $t_h/N$) with a gradient
modulation {$U=\prod_{k=0}^{N-1} \lb e^{-iH_z\tau k}U_0e^{iH_z\tau
    k}\rb$} that weights the couplings by a factor ${\cal G}_{ij}$,
with \begin{equation} \mathcal G_{ij} =\sum_{k=0}^{N-1}e^{ik\tau\xd\xo_{ij}} =
e^{i(N-1)\tau\xd\xo_{ij}/2}\frac{\sin(N\tau\xd\xo_{ij}/2)}{\sin(\tau\xd\xo_{ij}
  /2)}.  \end{equation}

We now make these ideas more concrete by considering a specific
example, the engineering of an Hamiltonian allowing perfect QIT in
mixed-state spin chains, thus enabling room temperature quantum
communication~\cite{Cappellaro11,Yao11,Ajoy12, Ajoy12x}.  For lossless transport, the simplest engineered
$n$-spin chain consists of only nearest-neighbor (NN) couplings that
vary parabolically along the chain,
$d_j=d\sqrt{j(n-j)}$~\cite{Christandl04,Albanese04}.  This ensures
perfect transport in a time $T=\pi/(2d)$.  However, manufacturing
chains with this precise coupling topology is a hard engineering
challenge due to fabrication constraints and the intrinsic presence of
long-range interactions~\cite{Strauch08, Ajoy12b}.  On the other hand,
\I{regular}, lattice-based spin networks (see~\zfr{network}) are found
ubiquitously in nature.  Our method can then be used to dynamically
engineer the optimal Hamiltonian starting from a complex spin network.
\paragraph{Filtered engineering for QIT --} 
Consider a dipolarly coupled spin network with Hamiltonian \begin{equation}
  \ham=\ham_{\R{int}}+\ham_z=\sum_{ij}b_{ij}(3S^z_iS^z_j-S_i\cdot
  S_j)+\sum_j\xo_jS_j^z, \end{equation} where the spatial \I{tagging} of the
  frequencies is achieved by applying an appropriate magnetic field
  gradient.  The target Hamiltonian for QIT in a $n$-spin chain is
  $H_d=\sum_{j=1}^{n-1} d_j(S_j^xS_{j+1}^x-S_j^yS_{j+1}^y)$. We
  consider this interaction, instead of the most commonly used
  isotropic XY Hamiltonian, since it drives the same transport
  evolution~\cite{Cappellaro07l,Ajoy12} and the double-quantum
  Hamiltonian $\ham_{\rm DQ} =\sum_{i<j} b_{ij} (S_{i}^xS_j^x -
  S_{i}^yS_j^y)$ can be obtained from the secular dipolar Hamiltonian
  via a well-known multiple-pulse sequence~\cite{Baum85,SOM}.  The
  sequence cancels the term $\ham_z$ and, importantly, allows
  time-reversal of the evolution by a simple phase shift of the
  pulses. We can further achieve evolution under the field gradient
only, $\ham_z$, by using homonuclear decoupling sequences, such as
WAHUHA~\cite{Waugh68,Haeberlen76} or magic echo
trains~\cite{Boutis03}. Alternatively, homonuclear decoupling can be
avoided by shifting the DQ sequence off-resonance during the free
evolution period~\cite{SOM}.

\begin{figure}[t]
 \centering
\includegraphics[width=0.5\textwidth]{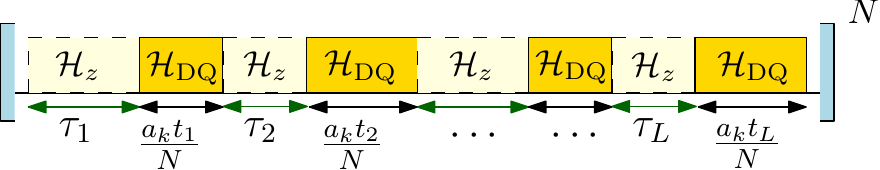}
\caption{A filtered engineering control sequence, consisting of alternating blocks
of (\I{free}) evolution under the gradient ${\cal H}_z$ for times $\tau_j$ and DQ (\I{mixing}) evolution ${\cal H}_{\R{DQ}}$ of duration $t_j/N$. 
The blocks are applied left to right, and the cycle is repeated $N$ times. The sequence can be \I{apodized} by an appropriate choice of coefficients $a_k$, where $k$ refers to the cycle number.
}
\zfl{sequence}
\end{figure}
A control sequence (see \zfr{sequence}) achieving filtered Hamiltonian engineering consists
of alternating blocks of \textit{free} evolution ($\B{\qt_{z}}$) under ${\cal H}_z$
and double-quantum excitation ${\cal H}_{\rm DQ}$ (\textit{mixing}
periods $\B{t_{m}}$). The mixing period might be further modulated by coefficients $a_k$ for improved robustness, as we explain in the following. 
We analyze this pulse sequence using average Hamiltonian theory~\cite{Haeberlen68,Haeberlen76}, setting for now $a_k=1\:\forall k$.  Consider for simplicity a sequence with only two mixing and free
evolution blocks.  Then, setting $U_z(\tau)\!=\!\exp({-}i\qt{\ham}_z)$
and $U_{\R{DQ}}(t)\!=\!\exp({-}it{\ham}_{\R{DQ}})$, the propagator
corresponding to $N$ cycles is $U_N
=\big[U_{z}(\tau_1)U_{\R{DQ}}\lb\frac{t_{1}}{N}\rb
U_{z}(\tau_2)U_{\R{DQ}}\lb\frac{t_{2}}{N}\rb\big]^N$, which can be
rewritten as \bea
U_N&\!=\!&\big[U_z(\tau_1)U_{\R{DQ}}\textstyle\lb\frac{t_{1}}{N}\rb
U^{\dag}_z(\tau_1)\big]
\zt\big[U_z{(\tau)}U_{\R{DQ}}\textstyle\lb\frac{t_{2}}{N}\rb
U^{\dag}_z{(\tau)}\big]\non\\
&\zt&{\cdots\zt\big[U_z{(N{\tau})}U_{\R{DQ}}\textstyle\lb\frac{t_{2}}{N}\rb
  U^{\dag}_z{(N{\tau})}\big]},\non \eea
where ${\tau}\!=\!\tau_1+\tau_2$. 
Now, $U_z(\tau)U_{\R{DQ}}({t})U^{\dag}_z(\tau)\!=\!e^{-itH_{m}(\tau)}$, where 
$H_{m}(\tau)= \sum_{i<j}b_{ij}(S_{i}^+S_j^+e^{i{\tau}\xd_{ij}}
+S_{i}^-S_j^-e^{-i\tau\xd_{ij}})$ is the \I{toggling} frame Hamiltonian with
$\xd_{ij}=\xo_{i} + \xo_j$. 
Employing the Suzuki-Trotter approximation~\cite{Suzuki90}, the propagator $U_N$ is equivalent to evolution under the {average} Hamiltonian,
\begin{equation}
\ov{H} = \sum_{i<j}\frac{b_{ij}}NS_{i}^+S_j^+\lb
{t_{1}}e^{i\tau_{1}\xd_{ij}} +
{t_{2}}e^{i(\tau_{1}+\tau_{2})\xd_{ij}}\rb{\cal G}_{ij} + \R{h.c.}
\zl{Hav}
\end{equation}
with ${\cal G}_{ij} =
e^{i(N-1)\qt\xd_{ij}/2}\frac{\sin(N\qt\xd_{ij}/2)}{\sin(\qt\xd_{ij}/2)}$. 

In general, for a sequence  of
free times $\B{\qt_z}\!=\!\{\qt_{1},\dots,\qt_{L}\}$ and  mixing
times
$\B{t_m}\!=\!\{t_{1},\dots,t_{L}\}$, the average Hamiltonian is
$\ov{H} = \sum_{i<j}S_{i}^+S_j^+F_{ij}(\B{\qt_z},\B{t_m}){\cal G}_{ij}(\qt) +
\R{h.c.}$, where $\qt=\sum_{j=1}^{L}\qt_{j}$ and we define the \I{weighting}
function,
\begin{equation}
F_{ij}(\B{\qt_z},\B{t_m})=\frac{b_{ij}}N\sum_k
t_k\exp\left(i\xd_{ij}\textstyle\sum_{h=1}^k\tau_h\right).
\end{equation}

The grating ${\cal G}_{ij}$ forms a sharp filter with maxima at $\qt\xd_{ij} =
2m\pi$. We assume a static linear 1D-magnetic field gradient is applied along a selected
chain of spins in the larger network, such that the $j^{\R{th}}$ spin frequency  is
$\xo_j=j\xo-\omega_0$, where $\xo_0$ is the excitation frequency.  
Each spin pair term of the DQ Hamiltonian acquires a spatial phase tag under the gradient field:  if $\xo\qt = \pi$ and $2\xo_0\qt = 3\pi -2m\pi$ the NN
couplings are preserved, while the NNN couplings lie at the minima of the grating,
and are decoupled (see \zfr{filter}). Other non-NN, off-chain couplings (see
\zfr{network}), lie at the grating side-lobes, and have greatly reduced
amplitudes for large $N$.
 
\begin{figure}[t]
 \centering
\includegraphics[width=0.4\textwidth]{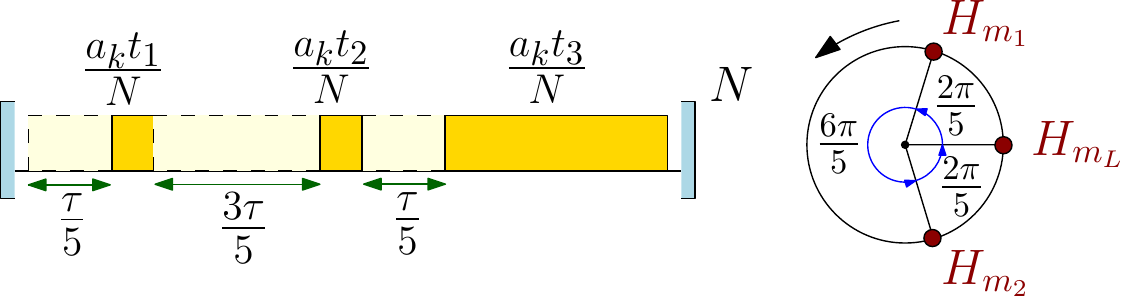}
\caption{Left: Example of control sequence engineering a  $n=5$ spin chain from a complex network such as in \zfr{network}. Here $\qt=\pi/\xo$, and
explicit values of $\B{t_m}=\{t_1,t_2,\cdots\}$ are in~\cite{SOM}. Right: 
Phasor representation of  Hamiltonian engineering. We show the phases $\phi_j=b_{i,i+1} t_j$  acquired by the $S_i^+S_{i+1}^+$ term of the toggling frame Hamiltonians
$H_{m_j}$ on a unit circle
(see also~\cite{SOM}).}
\zfl{sequence2}
\end{figure}

Following the filter, the weighting function  $F_{j(j+1)}$ can be constructed to
yield an engineered DQ Hamiltonian with the ideal coupling strengths required
for perfect transport. 
Given the toggling frame Hamiltonians, $H_{m}(\tau_k)$, we have a set of $2n$
equations (for an $n$-spin chain), 
\begin{equation}
\begin{array}{ll}
\sum_{h=1}^L\sin\!\left[\omega(2i+1){\sum_{k=1}^{h}\tau_k}\right] t_h
b_{i,i+1}=0,&\quad\forall i\vspace{10pt}\\
\sum_{h=1}^L\cos\!\left[\omega(2i+1){\sum_{k=1}^{h}\tau_k}\right] t_h
b_{i,i+1}\propto d_i,&\quad\forall i,\end{array}
\label{eq:SinCos}
\end{equation}
with $2L$ unknowns for $L$ time steps. The number of conditions (and thus of
time steps) can be reduced by exploiting symmetry properties. For example, by
imposing a gradient symmetric with respect to the center of the chain, it could
be possible to automatically satisfy most of the conditions in
Eq.~(\ref{eq:SinCos}) and only $L=\lceil n/2\rceil$ time steps would be required.
Unfortunately this solution is practical only for some chain lengths~\cite{SOM}; we thus
focus on a suboptimal, but simpler solution.
Consider for example an odd $n$-spin chain.  To enforce the mirror symmetry of the couplings
$d_j$ and ensure that the average Hamiltonian remains in DQ
form, we imposes mirror symmetry to the times, $t_{j} = t_{{L-j}}$, while the gradient evolution intervals are $\tau_{j}/\qt = 3/n$ for
$j=(L+1)/2$ and $\tau_{j}/\qt = 1/n$ otherwise (see \zfr{sequence2}). 
This choice  yields a simple \I{linear} system of  equations for  $L=n-2$
mixing periods $\B{t_m}$,
\begin{equation}
F_{j(j+1)}{\cal G}_{j(j+1)} = \sum_k t_k\cos\left(\frac{2j\pi k}{n}\right)=
d\sqrt{j(n-j})
\end{equation}
An intuitive phasor representation~\cite{Vinogradov99} of how the
evolution periods exploit the symmetries is presented in \zfr{sequence2} and in~\cite{SOM}. Analogous solutions can be derived for even spin chains.
\begin{figure}[t]
 \centering
\includegraphics[width=0.32\textwidth]{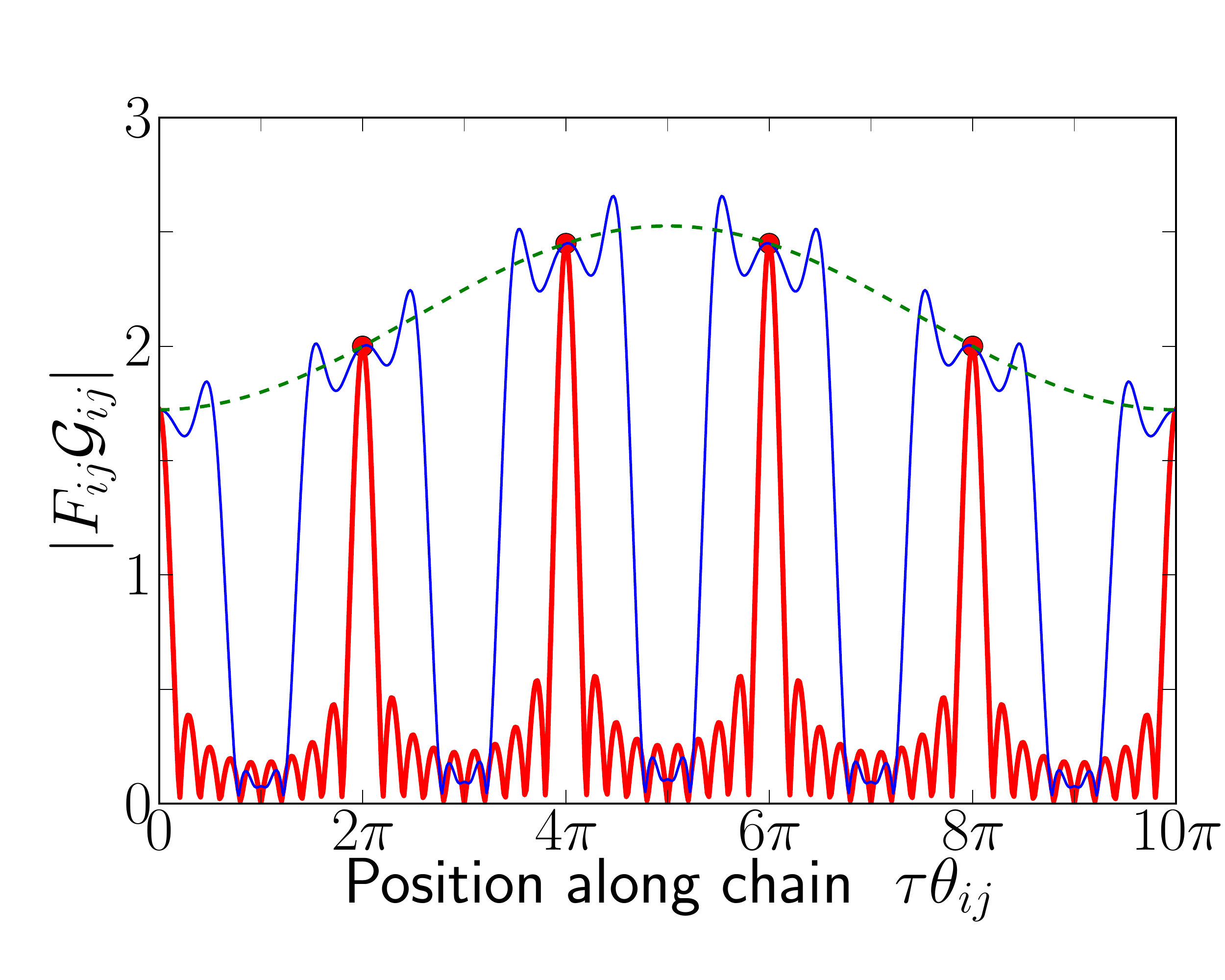}
\caption{Engineering filter function  $|F_{ij}{\cal G}_{ij}|$ for a 5-spin chain. 
A single cycle creates the weighting function $F_{ij}$ (dashed line), which is transformed to sharp (red) peaks at the ideal engineered couplings (red circles) as the number of cycles increases (here $N=10$). The peak widths can be altered by suitable apodization, for example the blue line refers to sinc-apodization with $a_k=\sin(W(k-N/2))/(W(k-N/2))$ with $W=(\pi+1)/2$ in \zfr{sequence}, and normalized such that $\sum a_k =N$.
}
\zfl{filter}
\end{figure}

The tuning action of $F_{j(j+1)}{\cal G}_{j(j+1)}$ is remarkably
rapid, achieving perfect transport fidelity in just a few cycles. 
Increasing the number of cycles  reduces the error in the Trotter expansion
improving $F_{j(j+1)}$ (\zfr{chain}.a) as well as improving the
filter selectivity of ${\cal G}_{j(j+1)}$ (\zfr{chain}.b). The peak width of the Bragg grating  decreases with the number of cycles as $2\pi/N$~\cite{Ajoy11}, improving the grating selectivity linearly with $N$~\cite{SOM}. 
As shown  in \zfr{chain}, about $n$ cycles are required for almost perfect decoupling of the unwanted interactions 
(fidelity $>0.95$). 

The highly selective grating also avoids the need for the chain to be isolated and for the surrounding network to have a regular structure: 
Any spin lying  between two NN of the chain can be effectively decoupled, as shown in \zfr{disorder}. However, in the presence of disorder in the couplings of the chain spins one needs to compromise between broader grating peaks (via small $N$) and poorer
decoupling of unwanted interactions.

To improve the robustness of our scheme to disorder and decoherence, we can further modulate the mixing times $t_{\vec{m}}$ in \zfr{sequence} by coefficients $a_k$; this imposes an apodization of the grating function as  ${\cal G}_{ij}=\sum a_ke^{ik\qt\xd_{ij}}$. 
Until now, our Hamiltonian engineering method has relied on  precise phase relationships between different couplings due to evolution under the gradient; however, even in the case of a simple dephasing noise interaction, these phase relationships are no longer exact. 
Apodization can counter dephasing and disorder -- the grating peaks can be made wider by $W$ (\zfr{filter}) and any coupling that is in phase to within $W$ can still be engineered robustly (see \zfr{apod}) at the expenses of a poorer  decoupling efficiency of long-range couplings.   Apodization may have other applications: for instance, it could be used to engineer \I{non-linear} spin chains in lattices~(see \cite{SOM}).

Quite generally, the apodized grating-based filter  could  be used to {select} any regular array of spins from a complex network -- allowing a wide
applicability of our method to many natural spin networks and crystal lattices. In ~\cite{SOM} we present a simple example employing the honeycomb lattice.
\paragraph{Approximation validity --}
The control sequence is designed to engineer  the average Hamiltonian $\ov{H}$ only to first order. 
Higher order terms yield errors scaling as ${\cal O}(t_{k}t_{k+1}/N^2)$, as arising from the Trotter expansion \cite{Suzuki90}.
Consider e.g. the propagator for a 5-spin chain,
\begin{equation}
U_N=[e^{-i\frac{t_{1}}{N}H_m(\tau_1)}e^{-i\frac{t_{1}}{N}
H_m(\tau_2)}e^{-i\frac{t_{L}}{N} {\cal H}_{\R{DQ}}}]^N,
\end{equation}
where $\tau_1=\pi/(5\omega)$ and $\tau_2=2\pi/(5\xo)$.
This yields the desired $\ov{H}$ with an error ${\cal O}(t_{1}^2/N^2)$ for
the first product, and ${\cal O}(2t_{1}t_{L}/N^2)$ for the second.
Increasing $N$ improves the approximation, at the expense of larger overhead
times (\zfr{chain}). 
In addition, the chosen  scheme achieves remarkably good fidelities even for
small $N$, since by construction, $t_j\ll t_L \approx T/N$. 
In essence, the system evolves under the unmodulated DQ Hamiltonian during $t_{L}$, yielding the \I{average} coupling strength, while the
$t_{j}$ periods apply small corrections required to reach the ideal couplings. 

Symmetrizing the control sequence would lead to a more accurate average Hamiltonian  because of  vanishing  higher orders~\cite{Levitt08}. 
However, this comes at the cost of longer overhead times $\B{\qt_{z}}$, thus using a larger number of the unsymmetrized sequence is often a better strategy. 
\begin{figure}[t]
 \centering
\includegraphics[width=0.45\textwidth]{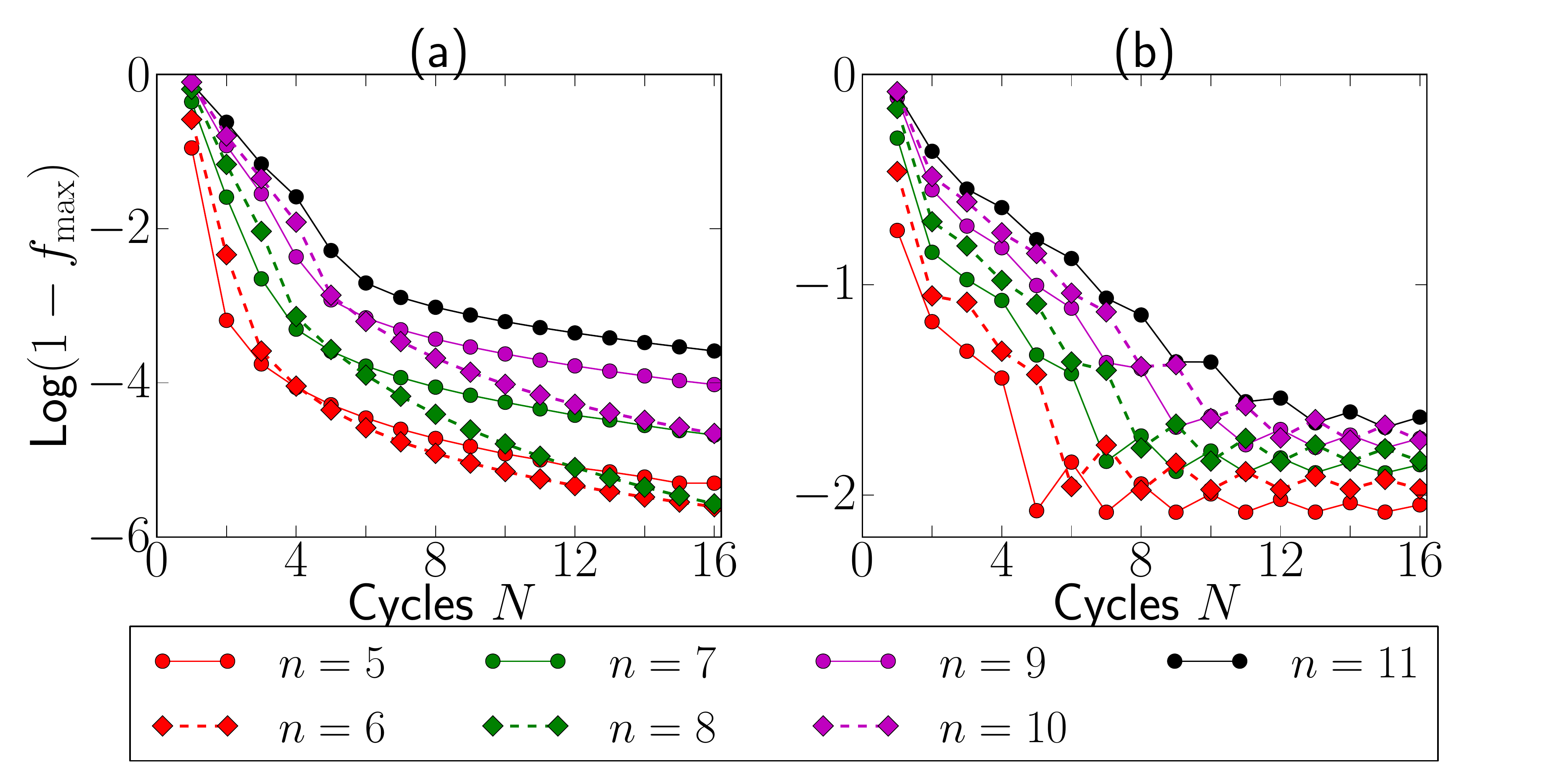}
\caption{Minimum transport infidelity obtained by filtered engineering,  as a function of cycle number $N$ for a $n$-spin  dipolar chain with (a) NN couplings only  and (b) all couplings included.  Here we calculated the fidelity
$f~=~\tr{US_1^zU^\dag S_n^z}/2^n$. Almost perfect fidelity is
achieved even for long chains with just a few filter cycles.
}
\zfl{chain}
\end{figure}

\paragraph{Experimental viability --}We now consider a possible
experimental implementation of the filtered engineering sequence and
show that high fidelity QIT at room temperature is achievable with
current technology.

We assume that the spin network is based on a physical lattice of NN
separation $r_0$, yielding a NN coupling strength
$b=\frac{\mu_0\hbar}{4\pi}\frac{\Gamma^2}{r_0^3}$. If an ideal
$n$-spin chain could be fabricated in this lattice with maximum
coupling strength $b$, the transport time would be $T_\R {id} =
\frac{n\pi}{8b}$~\cite{Cappellaro11}.  Alternatively, perfect state
transfer could be ensured in the weak-coupling
regime~\cite{Ajoy12x,Yao11}, with a transport time $T_\R
{weak} = \frac{\Gamma\pi}{b}$, where $\Gamma\gg1$ ensures that the
end-spins are weakly coupled to the bulk-spins.  We compare $T_\R
{id}$ and $T_\R{weak}$ to the time required for $N$ cycles of the
engineering sequence, $T_\R{eng}$.  Since $t_{L}\gg t_{j}$, to a good
approximation the total mixing time is
$Nt_L\lesssim\sum_j\sqrt{j(n-j)}/n\approx \pi n/8$.  Adding the
overhead time $N\tau$, which depends on the available gradient
strength as $\qt=\pi/\xo$, we have, 
$ T_\R {eng} = \frac{n\pi^2}{16b} + \frac{N\pi}{\xo}$. Since we can
take $\Gamma\!\approx\!n$ for the weak regime~\cite{Yao11} and
$N\!\approx\!n$ for filtered engineering, a gradient larger than the
NN coupling strength would allow for faster transport in the latter
case.

For concreteness, consider a crystal of fluorapatite (FAp)
[Ca$_5$(PO$_4$)$_3$F] ~\cite{Oishi94} that has been studied for
quantum transport \cite{Cappellaro07l}. The $^{19}$F nuclear spins
form parallel linear chains along the $c$-axis, with intra-chain
spacing $r_0=0.3442$nm ($b=1.289$kHz), while the inter-chain coupling
is about $40$ times weaker.  {States to be transported could be
  prepared on the ends of these chains~\cite{Kaur12}.}  Maxwell field
coils~\cite{Zhang98} can generate sufficient gradient strengths: for
instance, Ref.~\cite{Rumala06} demonstrated a gradient of
$5.588~\zt~10^8$G/m over a $1 \R{mm}^3$ region, corresponding to
$\xo=0.7705$kHz. Far stronger gradients are routinely used in magnetic
resonance force microscopy (MRFM)~\cite{Rugar04}; for example,
dysprosium magnetic tips~\cite{Mamin12} yield gradients of 60G/nm,
{linear} over distances exceeding $30$nm, yielding $\xo=82.73$kHz.
Setting $\xo=25$kHz would allow $\pi/2$-pulse widths of about
$0.5\mu$s to have sufficient bandwidth to control chains exceeding
$n=50$ spins.  Homonuclear decoupling sequences~\cite{Haeberlen76,
  Boutis03} can increase the coherence time up to $T_{\R{eng}}$.  For
FAp, evolution under the DQ Hamiltonian has been shown to last for
about $1.5$ms \cite{Cappellaro07l}; we anticipate that decoupling
during the $U_z$ periods could increase this easily to
$15$ms~\cite{Boutis03}. While pulse errors might limit the performance
of Hamiltonian engineering, there exist several methods to reduce
these errors~\cite{Levitt86}.  With $\xo=25$kHz, and $N=30$ cycles,
nearly lossless transport should be possible for a 25-spin chain.

\begin{figure}[t]
 \centering
\includegraphics[width=0.45\textwidth]{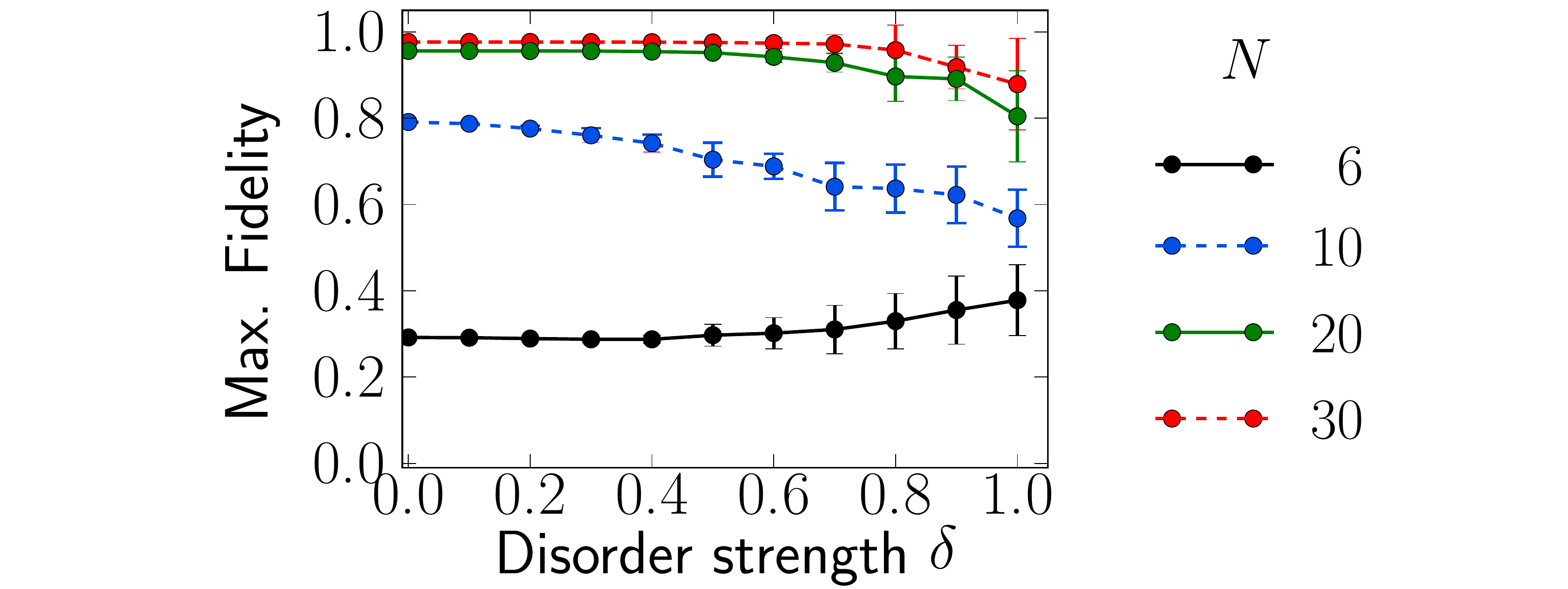}
\caption{
time for $N=30$.  (b) 
Variation of maximum fidelity with disorder in the network surrounding the spin chain in \zfr{network}. 
The spins are displaced by $\xd\cdot r$,
where $r$ is a  random
number {uniformly distributed} on $[-x/2,x/2]$ (averaged over 30 realizations)
with $x$  the NN chain spin separation.}
\zfl{disorder}
\end{figure}

\begin{figure}[t]
 \centering
\includegraphics[width=0.45\textwidth]{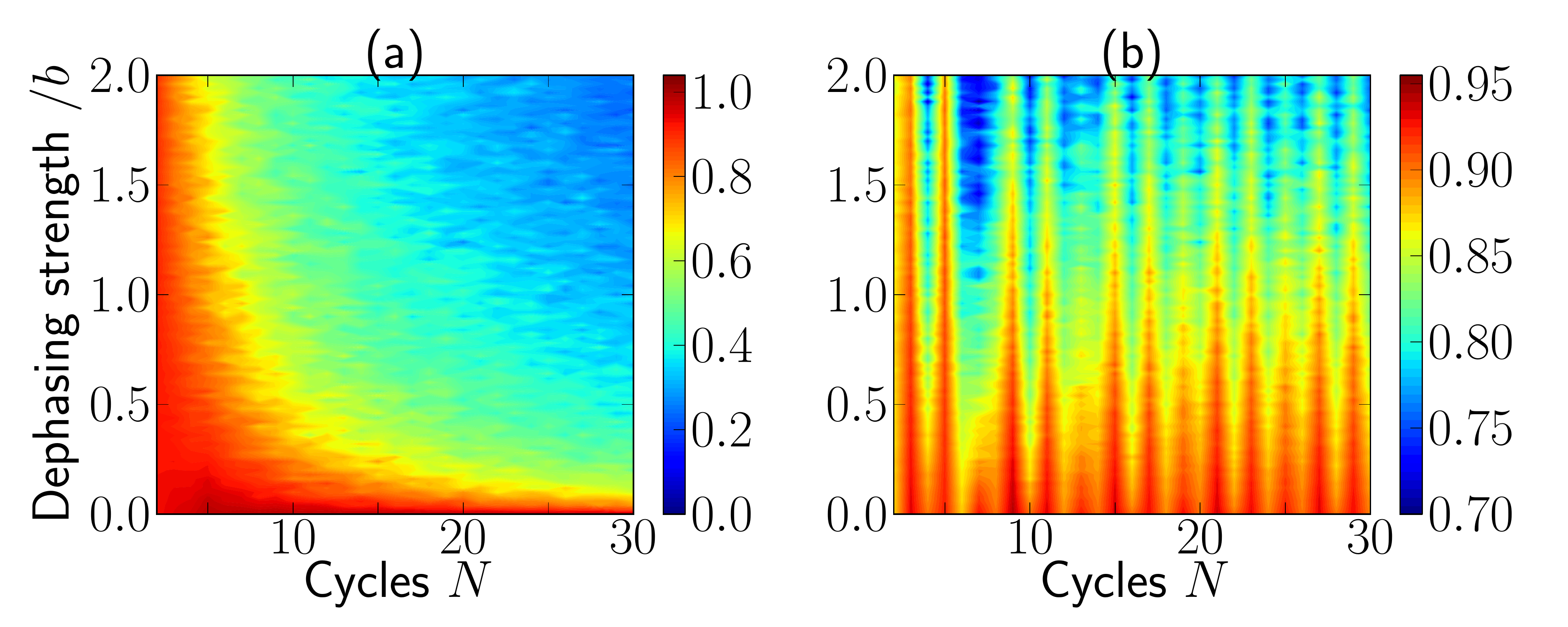}
\caption{Maximum transport fidelity $f$ for a  $n=5$ spin  chain with dipolar couplings ($b$ is the NN coupling strength).  The spins are subject to  dephasing noise,  modeled by an Ornstein-Uhlenbeck process of correlation time $\qt_c = 2/b$ and  strength $2b$, averaged over 100 realizations .  (a)
No apodization. (b) With sinc-apodization ($W=(\pi+1)/2$ as in
\zfr{filter}). }
\zfl{apod}
\end{figure}

Filtered Hamiltonian engineering
 could as well be implemented in other
physical systems, such as trapped ions~\cite{Blatt12} or optical
lattices~\cite{Simon11}. For instance, Rydberg atoms in optical
lattices~\cite{Saffman10,Weimer10} could enable simulations at low
temperature by the filtered engineering method, thanks to the
availability of long-range couplings and the ability to tune the
lattice to create gradients. The scheme could also be extended to more complex 2D and 3D lattices (\cite{SOM}).
\paragraph{Conclusion --}We have described a general method for
quantum simulation which does not require local control, but relies on
the construction of time domain filter and weighting functions via
evolution under a gradient field. The method was applied to engineer
spin chains for perfect transport, isolating them from a large,
complex network.  We showed that robust and high fidelity quantum
transport can be driven in these engineered networks, with only
experimental feasible control.

\bibliographystyle{apsrev4-1}
\bibliography{Biblio_001}

\newpage 

\onecolumngrid

\section*{Supplementary Information}

\setcounter{figure}{0}
\setcounter{equation}{0}
\setcounter{section}{0}

\section{Operator engineering} 
An important ingredient of any Hamiltonian engineering scheme is the ability to obtain the Hamiltonian operator form desired, before engineering the coupling strengths.
Nuclear Magnetic Resonance has a long tradition of control sequences able to modify the naturally occurring Hamiltonians into desired operators; the most prominent application is in the refocusing of unwanted interactions, which has been further developed and is now a common technique in quantum information under the name of ``dynamical decoupling''~\cite{Viola99b}. 
Here we briefly review  these decoupling sequences and extend them to more general two-body Hamiltonians as well as to the task of creating a desired operator.

We consider a general Hamiltonian for 2 spin-$\frac12$ particles:
\begin{equation}
H=\vec{\omega}_1\cdot\vec{\sigma}_1+\vec{\omega}_1\cdot\vec{\sigma}_1+\vec{\sigma}_1\cdot\mathbf{d}\cdot\vec{\sigma}_2,
\end{equation}
which can be rewritten in terms of spherical tensors
$T_{l,m}$~\cite{rose} (see \ztr{tensors}):
 \begin{equation}
	H=\sum_{l,m} (-1)^mA_{l,-m} T_{l,m}
\end{equation}
where the coefficients $A_{l,-m}$ depend on the type of spin-spin interaction
and the external field. This notation is useful when considering rotations of
the Hamiltonian. We consider only collective rotations -- {that is,
rotations applied to all the spins together, which are easily achievable
experimentally. Note that they also conserve the rank $l$~\cite{Haeberlen76}.}

\begin{table}[b]
	\centering
		\begin{tabular}{ll}
	\hline\hline\vspace{-6pt}\\
	$ \openone $ & $ T_{00}\!=\!(\sigma^a_x\sigma^b_x\!+\!\sigma^a_y\sigma^b_y\!+\!\sigma^a_z\sigma^b_z)/\sqrt{3}  $ \\ 
	$T^a_{10}\!=\!\sigma^a_z/2 $ & $ T^b_{10}\!=\!\sigma^b_Z/2 $ \\ 
$T^a_{11}\!=\!\sigma^a_+/\sqrt{2} $ & $T^a_{1-1}\!=\!\sigma^a_-/\sqrt{2} $\\ 
$ T^b_{11}\!=\!\sigma^b_+/\sqrt{2} $  & $ T^b_{1-1}\!=\!\sigma^b_-/\sqrt{2} $ \\ $
T_{11}\!=\!(\sigma^a_+\sigma_z^b\!-\!\sigma^a_z\sigma_+^b)/2 $ & $ T_{1-1}\!=\!(\sigma^a_-\sigma_z^b\!-\!\sigma^a_z\sigma_-^b)/2  $ \\ $
T_{10}\!=\!(\sigma^a_+\sigma_-^b\!-\!\sigma^a_-\sigma_+^b)/2  $ \quad\quad& $ T_{20}\!\!=\!\!(2\sigma_z\sigma_z\!-\!\sigma^a_x\sigma^b_x\!-\!\sigma^a_y\sigma^b_y)/\sqrt{6}$ \\ $
T_{21}\!=\!(\sigma^a_+\sigma_z^b+\sigma^a_z\sigma_+^b)/2 $ & $T_{2-1}\!=\! (\sigma^a_-\sigma_z^b+\sigma^a_z\sigma_-^b)/2 $ \\ $
T_{22}\!=\!\sigma^a_+\sigma^b_+/2 $ & $T_{2-2}\!=\!\sigma^a_-\sigma^b_-/2  $\vspace{4pt}\\\hline\hline
	\end{tabular}
	\caption{Spherical tensors for two spin-1/2 ($a$ and $b$)~\cite{rose}.
$\sigma_\alpha$ are the usual Pauli operators.}
	\ztl{tensors}
\end{table}

\begin{figure*}[t]
 \centering
\includegraphics[scale=0.8]{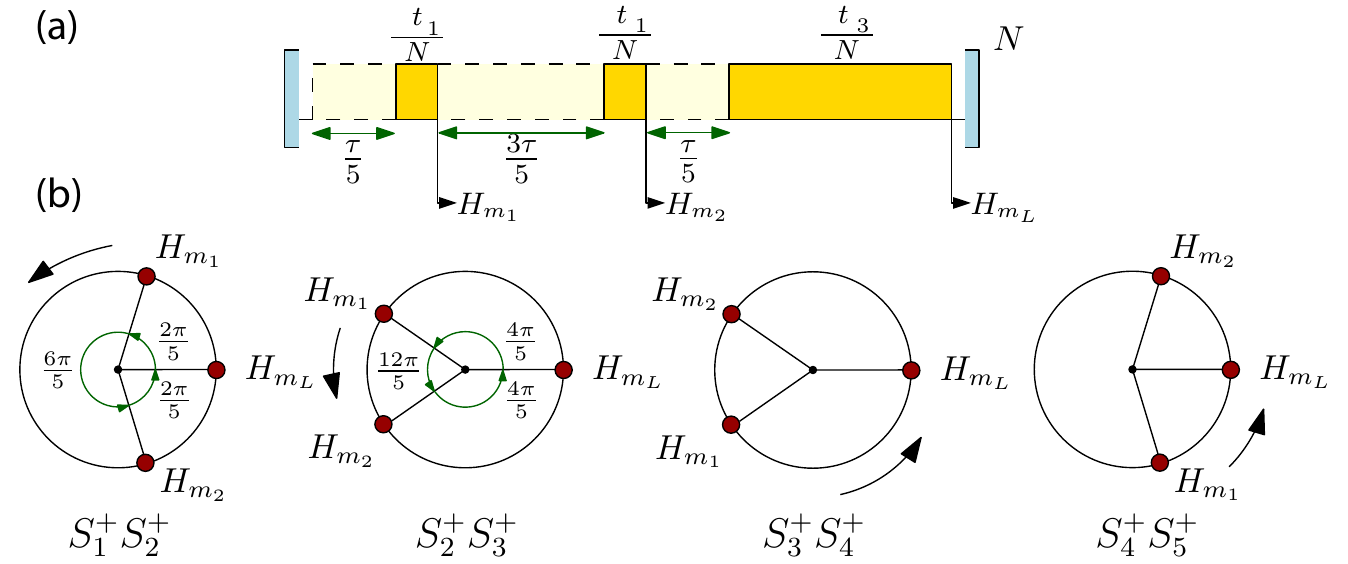}
\caption{Phasor representation of the Hamiltonian engineering sequence for $n=5$. The phase
acquired by the toggling frame Hamiltonians $H_{m_j}$ after each mixing period
in the sequence (a) is depicted on a unit circle for each of the NN terms
$S^+_{j}S^+_{(j+1)}$ in (b). The unit circle is traversed anti-clockwise from
$H_{m_1}$ to $H_{m_L}$. Mirror symmetry of the engineered couplings about the
center of the chain is ensured since $\{S^+_1S^+_2,S^+_4S^+_5\}$ and
$\{S^+_2S^+_3,S^+_3S^+_4\}$ have identical phasor representations. The DQ form
of the average Hamiltonian is ensured by forcing that the phasors for each of
the $S^+_{j}S^+_{(j+1)}$ terms have reflection symmetry about $2\pi$. }
\zfl{phase2}
\end{figure*}
\begin{figure*}[t]
 \centering
\includegraphics[scale=0.6]{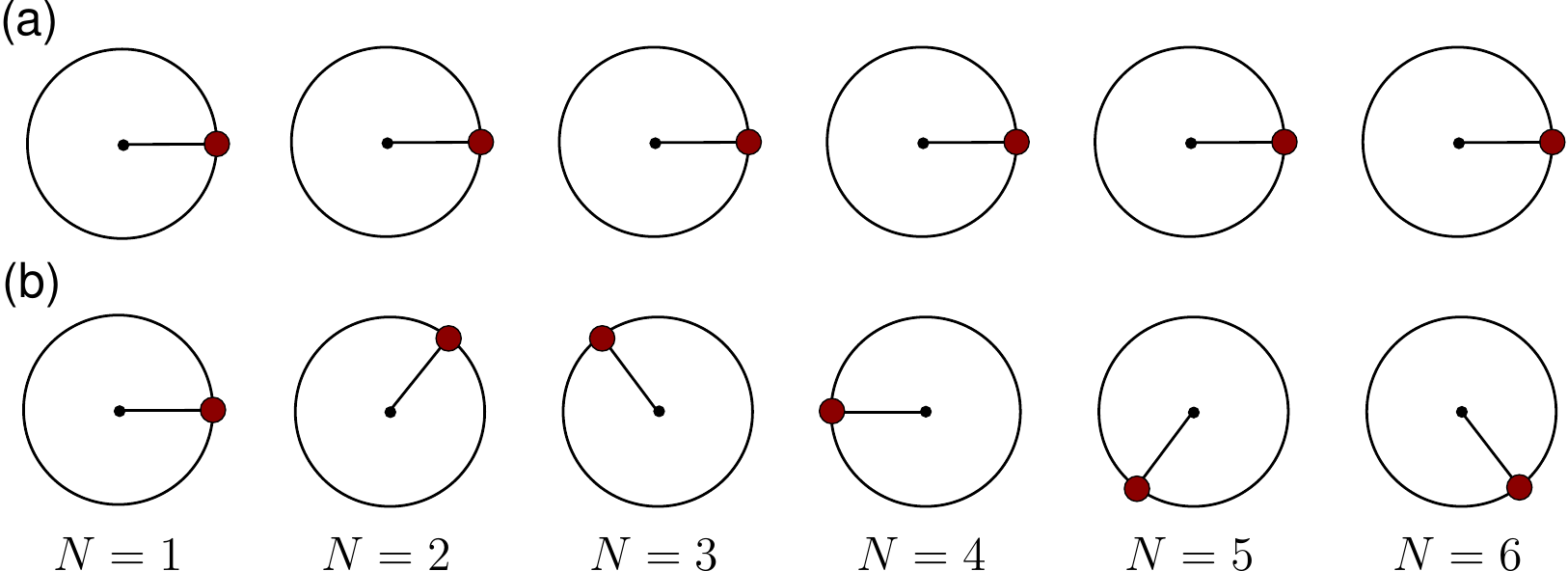}
\caption{Phasor representation of the grating's filtering action upon cycle iteration. (a) The effective phasor of a nearest-neighbor neighbor coupling traverses a $2\pi$ rotation with each cycle, and is perfectly refocussed (constructive interference). (b) However, the phasor corresponding to long range couplings do not complete a full rotation (the figure shows an example of a long range coupling for which $\xd_{ij}=\pi/3$). Hence cycle iteration leads to destructive interference and reduced average amplitudes of these couplings. }
\zfl{phase3}
\end{figure*}

\begin{table*}[t]
\begin{tabular}{|ccc|}\hline\hline
Chain Length &\ \ & Mixing periods $\B{t_m}$  \\\hline
$4$ &\ &$\{   -0.1340,   -0.1340,    1.7321\}$  \\
$5$ &\ &$\{-0.201,-0.201,2.1242\}$  \\
$6$ &\ &$\{ -0.2962, -0.0415,  -0.0415, -0.2962,  2.4907\}$\\
$7$ &\ &$\{-0.369,-0.0652,-0.0652,-0.369,2.8806\}$  \\
$8$ &\ &$\{ -0.4533,   -0.1024,   -0.0194,   -0.0194,   -0.1024,   -0.4533,    3.2594\}$\\
$9$ &\ &$\{-0.5269,-0.1298,-0.0316,-0.0316,-0.1298,-0.5269,3.6492\}$\\
$10$ &\
&$\{   -0.6065,   -0.1638,   -0.0517,   -0.0109 ,  -0.0109,   -0.0517,   -0.1638,   -0.6065,    4.0330\}$\\
$11$ &\
&$\{-0.6801,-0.1920,-0.0666,-0.0183,-0.0183,-0.0666,-0.1920,-0.6801,4.4231\}$\\
\hline
\end{tabular}
\caption{Mixing times $\B{t_m}$ in units of $1/d$ (as used in Fig. 5 and Fig. 6 of the main text). Note that $t_L\gg t_j$ for
$j\neq
L$. Negative times are easily achieved by a $\pi/2$ shift of the DQ sequence pulses, which implements a time-reversal version of the DQ Hamiltonian.}
\ztl{times}
\end{table*}
The goal of Hamiltonian engineering by multiple pulse sequence is to obtain a desired Hamiltonian from the naturally occurring one by piece-wise constant evolution under rotated versions of the natural Hamiltonian. We thus want to impose the condition
\beq
\sum_k R_k \ham_{nat} R_k^\dag = \ham_{tar},
\label{eq:HamEng1}
\eeq
where $R_k$ are collective rotations of all the spins, which achieves the desired operator to first order in a Trotter expansion.
Rotations can be described by the Wigner matrices $D^l_{m,n}(R_k)$ as:
\begin{equation}
	R_kT_{l,m}R_k^\dag=\sum_{n}(-1)^mD^l_{m,n}(R_k)T_{l,n}
\end{equation}
thus Eq.~(\ref{eq:HamEng1})  can be written using the spherical tensors as
\begin{equation}
	\sum_{k,n}\sum_{l,m}(-1)^mA^{nat}_{l,-m}D^l_{m,n}(R_k)T_{l,n}
= \sum_{l,m} (-1)^mA^{des}_{l,-m} T_{l,m},
\end{equation}
yielding the set of equations
\begin{equation}
	\sum_{k,m}(-1)^mA^{nat}_{l,-m}D^l_{m,n}(R_k)= (-1)^nA^{des}_{l,-n}. 
	\label{eq:HamEng}
\end{equation}
We note that since collective pulses cannot change the rank $l$, there are limitations to which Hamiltonians  can be engineered.
In particular,  $T_{00}$ commutes with collective rotations:  its contribution is thus a constant of the motion and, conversely, it cannot be introduced in the desired Hamiltonian if it is not present in the natural one. 
For example, an Ising Hamiltonian  $H_I=\sigma_z^{a}\sigma_z^{b}$ is expanded as $H_I=(T_{00}+\sqrt{2}T_{20})/\sqrt{3}$, so that only the second part can be modulated. Conversely, the secular dipolar Hamiltonian is given by $T_{20}$, thus it cannot produce an Hamiltonian containing $T_{00}$ (although it can be used the create the DQ-Hamiltonian, $\ham_{\R{DQ}}=T_{22}+T_{2,-2}$). 
Group theory methods can be used to help solving Eq. (\ref{eq:HamEng}) by
reducing the number of conditions using symmetries. For example, the DQ
Hamiltonian can be prepared from the secular dipolar Hamiltonian by using a
simple sequence consisting of two time intervals, $t_1=t_2/2$ with the
Hamiltonian rotated by $R_2=\left.\frac\pi2\right|_y$ in second time period, to
yield:
$T_{2,0}t_1+\left[\sqrt{\frac3{128}}(T_{2,2}+T_{2,-2})-{\frac12T_{2,0}}\right
] t_2\!\propto\!\ham_{DQ}$. Symmetrized versions of this simple sequence are
routinely used in NMR experiments~\cite{Baum85, Suter87}.

\begin{figure}[t]
\centering
\includegraphics[width=0.5\textwidth]{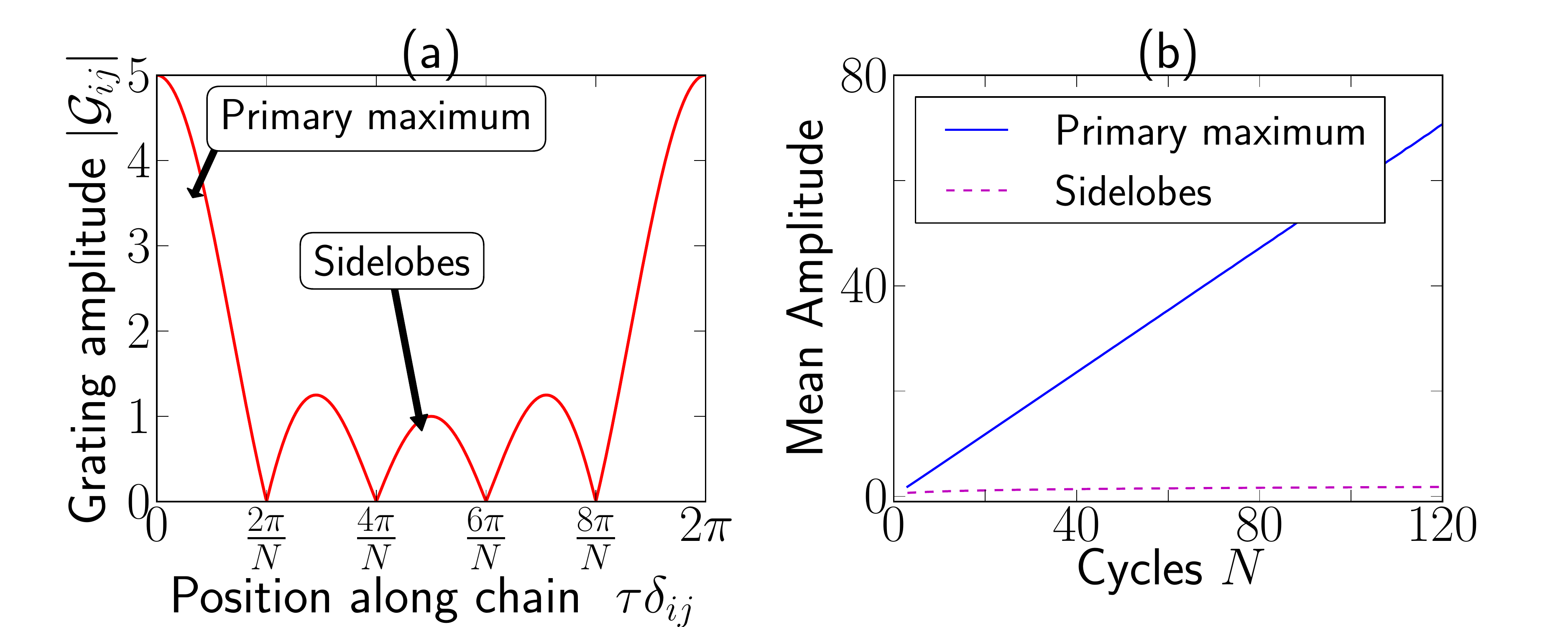}
\caption{(a) The amplitude of the dynamic Bragg grating, $|{\cal G}_{ij}|$ for
$N=5$ cycles. There are $N-1$ minima at $2j\pi/N$, and $N-2$ secondary maxima
(\I{sidelobes}). The width of the primary maximum is $2\pi/N$, and decreases
with the number of cycles making the grating more selective. (b) Right panel
characterizes the decoupling efficiency in terms of the mean amplitude of
grating $|{\cal G}_{ij}|$ in the primary maxima (i.e the range
$[0,\pi/N]\cup[2(N-1)\pi/N,2\pi/N]$), and the side lobes (the range $[\pi/N,
2(N-1)\pi/N]$). Increasing the number of cycles decreases the relative sideband
power, and hence increases the grating selectivity.} 
\zfl{filter-charac}
\end{figure}

\section{Phasor representation of the engineering sequence}
The filtered engineering sequence  presented in the main text has an intuitive geometric visualization in terms of phasors, 
which demonstrates the symmetries involved in the  QIT Hamiltonian and the subsequent choice of
 free evolution periods $\B{\qt_z}$. 
Let us consider for example the 5-spin sequence (\zfr{phase2}a), and the toggling frame Hamiltonians $H_{m_j}$ during each mixing period. 
The phases acquired by the Hamiltonians are represented on a unit circle for each of the $S_j^+S_{j+1}^+$ terms (\zfr{phase2}b). 
The phasor representation makes clear the symmetries involved in the $F_{ij}$ filter engineering. 
Since the ideal couplings ${d}_j$ are mirror symmetric about the center of the
chain, the construction of the filter $F_{j(j+1)}$ should match this symmetry.
This can be ensured in two ways.

We could select a gradient centered around the middle spin in the chain. 
Then $\omega_j=-\omega_{n-j}$, and the  set of equations in~(4) of the main text reduces to only $n-1$ equations (instead of $2(n-1)$) if $t_k=t_{L-k}$ and $\ham_m(\tau_k)=\ham_m(\tau_{L-k})^{\intercal}$. This second condition further ensures that the first set of equations in Eq. (4) are always satisfied (the condition is easily satisfied by setting $t^z_{k}=-t^z_{L-k}$ with $t^z_k=\tau_k-\tau_{k-1}$ and $t^z_1=\tau_1$). Thus we only have $\lceil (n-1)/2\rceil$ equations to be satisfied and correspondingly only $\lceil (n-1)/2\rceil$ mixing periods.  Although solutions can be found with this scheme, only for $n=4,5$ all the times are real. For larger spin chains it is possible to find all real solutions for $L=\lceil (n-1)/2\rceil+2$ mixing periods, although the system of equations quickly becomes intractable. 

A second strategy is to use a gradient with minimum at the first spin in the chain, $L=n-2$ mixing periods (for $n$ odd) and choose $\qt_{j}/\qt =3/n$ for $j=(L+1)/2$ and
$\qt_{j}/\qt = 1/n$, and $t_{j} = t_{{(L-j)}}$. 
Moreover,
since we want to retain the DQ form of the average Hamiltonian $\ov{H}$, the
toggling frame Hamiltonians $H_{m}(\tau_k)$ and $H_m(\tau_{L-k})$ are such
that their phasors have reflection symmetry about $2\pi$ for each
$S_j^+S_{j+1}^+$ term. This ensures that only the $(S_j^+S_{j+1}^+ +
S_j^-S_{j+1}^-)\cos[\qt\xd_{j(j+1)}]$ term survives, leading to an effective DQ
form (see \zfr{phase2}). 

The filtering action of the grating can also be visualized quite simply by the phasor diagrams in \zfr{phase3}: nearest-neighbor couplings undergo constructive interference upon cycle iteration and are retained, while long range couplings vanish on average.

\section{Characterizing grating selectivity}
Let us now characterize the selectivity of $N$-cycles of the Bragg grating
${\cal G}_{ij} =\sum_{k=0}^{N-1}e^{ik\qt\xd_{ij}}$. \zfr{filter-charac}(a) shows
the grating amplitude $|{\cal G}_{ij}|$ for $N=5$ cycles. In general, the width
of the primary maximum, $2\pi/N$, decreases with $N$ and this increases the
selectivity of the grating. Concurrently, the relative amplitude of the $N-2$
sidelobes decreases with $N$. Specifically, consider \zfr{filter-charac}(b),
which compares the mean grating amplitude in the primary maxima i.e. in the
range
$[0,\pi/N]\cup[2(N-1)\pi/N,2\pi/N]$, and the sidelobes in the range $[\pi/N,
2(N-1)\pi/N]$. From the properties of the Bragg grating, we find that the decoupling efficiency, characterized by the
relative amplitude of the amplitude in the primary maxima with respect to the
sidelobes, increases linearly with $N$.

\section{Construction of apodized sequences}

We  now describe more in the detail the apodized sequences we introduced in Fig.~7 of the main paper in order to make the Hamiltonian engineering more robust under decoherence. Apodized fiber Bragg gratings (FBGs) are widely used in photonics for suppression of sidelobe amplitudes (eg. \cite{Kashyap99b}). However, the novelty of our formalism is to demonstrate (i) that the same underlying principles can be applied in the \I{time domain} as opposed to the spatial construction in FBGs, (ii) that they have the extremely beneficial effect of making Hamiltonian engineering \I{robust} under environmental noise, and (iii) that they can be used to engineer \I{non-linear} crystal lattices where conventional comb-like gratings would fail.
\begin{figure}[h]
 \centering
\includegraphics[width=0.5\textwidth]{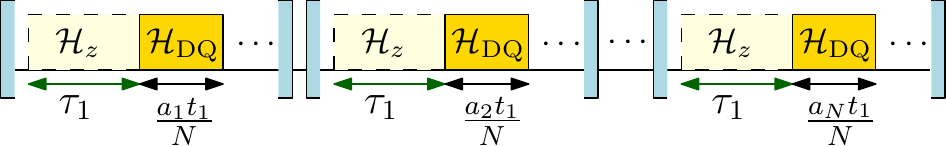}
\caption{Schematic of apodized Hamiltonian engineering sequences. Each cycle of the sequence is enclosed in blue brackets, and the sequence consists of $N$ cycles. The DQ evolution period in each cycle is weighted by the apodization coefficients $a_k$, where $k$ is the cycle number. Note that the net evolution period in this case remains almost identical to the case of non-apodized sequences. Sinc-apodization can be used to broaden the Bragg maxima, making the sequence more robust to dephasing; for this case we choose  $a_k=\sin(W(k-N/2))/(W(k-N/2))$, and normalize it such that $\sum a_k=N$. Here $W$ denotes the width of the maxima. 
}
\zfl{apod-sequence}
\end{figure}

The schematic of the apodized sequences are illustrated in \zfr{apod-sequence}, where $a_k$ denote the apodization coefficients with $k$ being the cycle number. In this case, the grating function gets modified as
\beq
 \mathcal G_{ij} =\sum_{k=0}^{N-1}a_k e^{ik\tau\xd_{ij}} = a_0 + a_1e^{i\tau\xd_{ij}} +  a_2e^{i2\tau\xd_{ij}} + \cdots +  a_{N-1}e^{i(N-1)\tau\xo_{ij}},
 \eeq
 which is just the N-point discrete Fourier transform of the apodization window function whose samples are $a_k$. Hence choosing the appropriate apodization coefficients $a_k$ allow one to engineer the grating. 
\begin{figure*}[h]
 \centering
\includegraphics[width=0.5\textwidth]{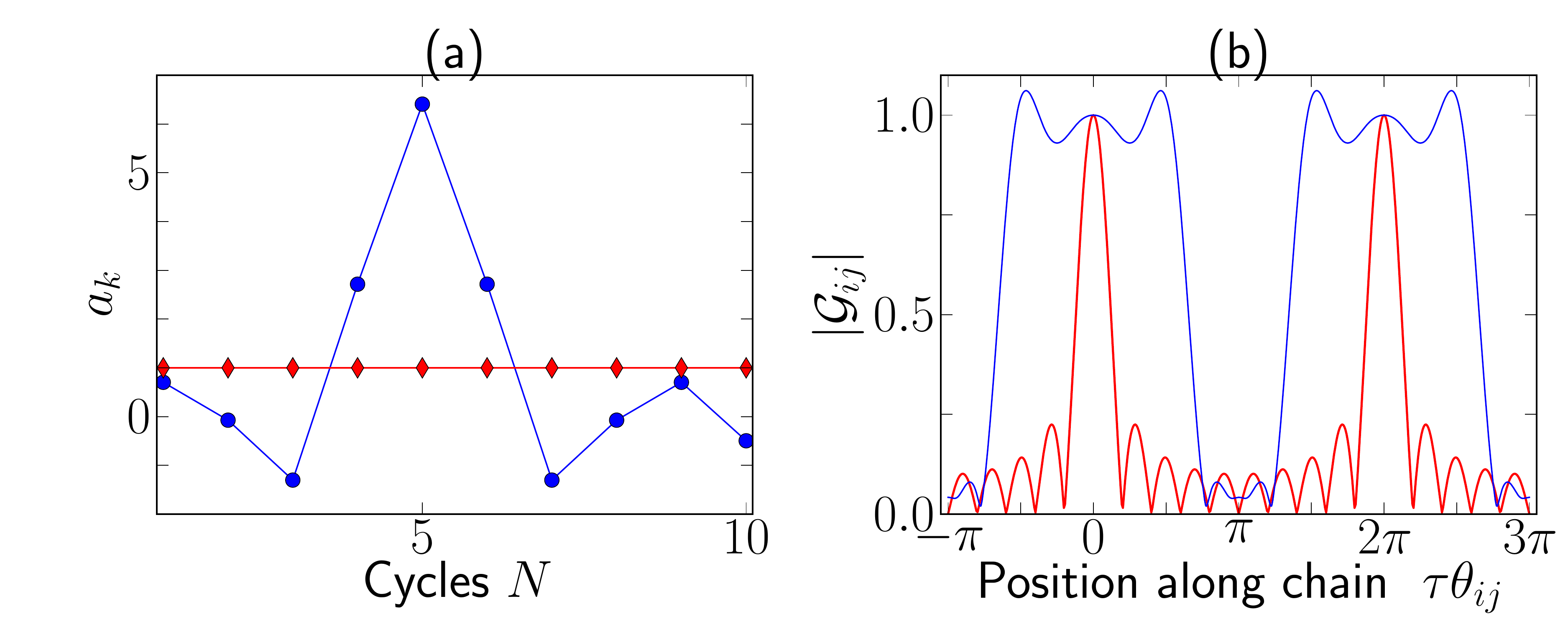}
\caption{(a) The blue circles show the apodization window $a_k$ upon sinc-apodization for $N=10$ cycles with $W=\pi/2 + 1/2$. The red diamonds show the case without apodization, i.e. $a_k=1$. Note that the DQ evolution period in the former case is about 1.65 times longer than without apodization; however this is only a marginal increase. (b) The grating function ${\cal G}_{ij}$ in case of sinc-apodization (blue) consists of broad square peaks of width $W$, compared to the sinc peaks in the non-apodized case (red).   }
\zfl{window}
\end{figure*}
For example, in the non-apodized case $a_k=1$, the window function is a rectangle, and the grating consists of sharp sinc-peaks (see \zfr{window}(a)). Under sinc-apodization, $a_k=\sin(W(k-N/2))/(W(k-N/2))$, the grating consists of square peaks of width $W$ (\zfr{window}(b)). Note that we normalize the window such that $\sum a_k=N$. This leads to a net DQ evolution period that is about 1.65 times longer than in the non-apodized case (for $N=10$); however, this is a marginal increase since the total length of the sequence depends more strongly on the free evolution period $N\pi/\xo$ (more precisely  $T_{eng} \approx \frac{n\pi^2}{16b} + \frac{N\pi}{\xo}$). In the apodized case, all couplings in $W$ satisfy the maxima condition and are refocussed. More importantly, the average sidelobe amplitude remains the same as that of the non-apodized case, leading to the same decoupling efficiency that grows linearly with $N$ (see \zfr{filter-charac}). Hence this leads to robustness under dephasing, since even couplings that have dephased by an amount $W/2$ are refocussed and amenable to Hamiltonian engineering. However, the mismatch to the exact engineered strengths due to the falling square profiles of the resulting $F_{ij} \mathcal G_{ij}$ functions leads to reduced fidelities in this case (see Fig. 7 of main paper).

\begin{figure*}[h]
  \centering
\subfigure[]{\includegraphics[width=0.5\textwidth]{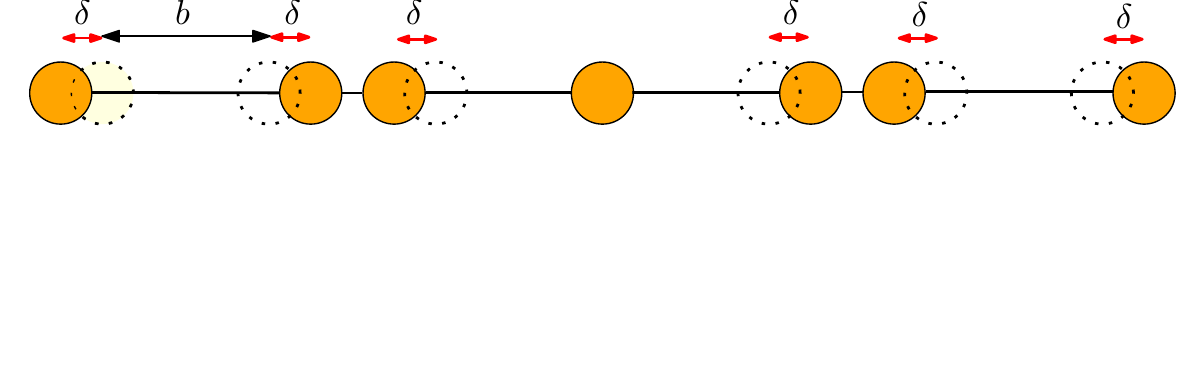}}
 \subfigure[]{\includegraphics[width=0.3\textwidth]{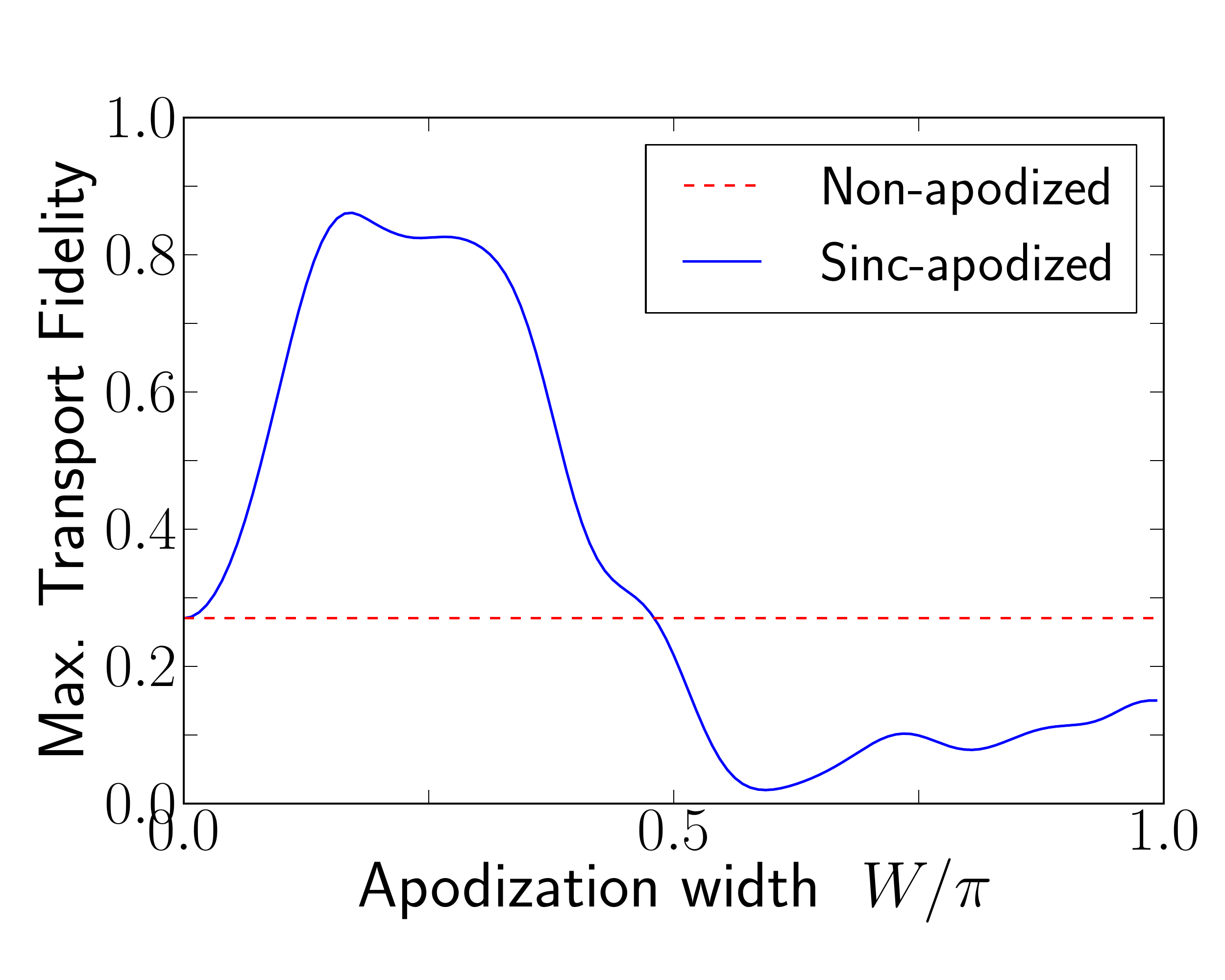}}
 \caption{Apodized sequences can also aid in the engineering of a non-linearly spaced array of spins. For instance, panel (a) shows a non-linearly spaced $n=7$ spin chain (orange) obtained by perturbing a regular chain (dotted), with perturbation strength $\xd/b = 0.15$. For simplicity we consider a mirror symmetric perturbation; however this is not necessary in general. The strengths of the couplings for perfect transport are calculated by modifying the usual parabolic strengths $d_{j(j+1)} = b\sqrt{n(n-j}$ by also considering the different dipolar strengths between neighboring spins. However, even when perfectly engineered, the sequence would not work efficiently since the spins are non-linearly spaced while the usual grating is linear. (b) Sinc- apodization provides a possible solution. The blue line shows the transport fidelity for different apodization widths $W$ considering all couplings included and $N=20$ cycles, and demonstrates high fidelity $>0.85$. For large $W$, the fidelity drops because certain long-range couplings now get refocussed. Without apodization, the fidelity is far lower (red dashed) especially for large $N$.}
\zfl{nonlinear-chain}
\end{figure*}
 In \zfr{apod2}, we study in more detail the case considered in Fig. 7 of the main paper of a $n=5$ spin chain embedded in a spin bath and under the influence of Ornstein-Uhlenbeck noise due to it. The results clearly show that apodization leads to more robust performance even under strong noise fields (for instance, strengths exceeding the nearest-neighbor coupling $b$, \zfr{apod2}(a)), and long application times of the sequence (\zfr{apod2}(b)).

\begin{figure}[h]
 \centering
\includegraphics[width=0.55\textwidth]{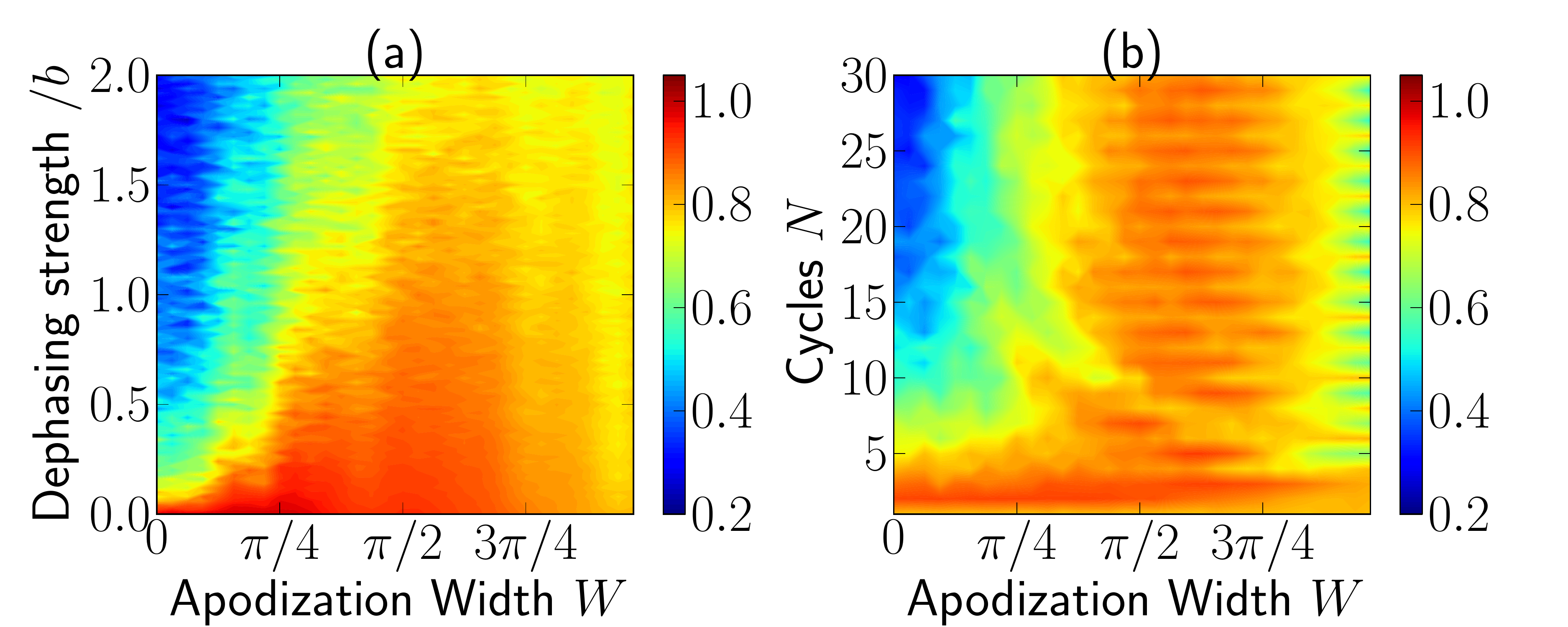}
\caption{Maximum transport fidelity $f$ for a fully coupled $n=5$ spin dipolar chain embedded in a spin bath. We consider that the bath produces a purely dephasing noise field on the chain, modelled by an Ornstein-Uhlenbeck (OU) process of correlation time $\qt_c = 2/b$, and  maximum dephasing strength $2b$, where $b$ is the NN coupling strength (see also Fig. 7 of the main paper). (a) Here we consider $N=20$ cycles, and study the variation of transport fidelity with sinc-apodization width $W$, averaged over 100 manifestations of the OU noise. When $W=0$ (no apodization), the fidelity rapidly decreases even under small noise fields. As $W$ increases, the broader Bragg maxima lead to more robust performance. There exists an optimal $W$ that maximizes fidelity -- in this case $W\app 5\pi/8$ produces fidelity exceeding 0.9 even for noise strengths $\app b$. (b) In the right panel, we study the variation of maximum fidelity under a fixed dephasing strength $=b$, for varying $W$ and number of cycles $N$. Without apodization,  increasing $N$ leads to sharper Bragg peaks and poor performance under noise. At the optimal $W\app 5\pi/8$, there is extremely robust performance even for large $N$, demonstrating that the sequence can be applied for long times without loss of performance. Note that the figure also shows that even cycles perform slightly better than when $N$ is odd, due to better symmetrization.}
\zfl{apod2}
\end{figure}

 The apodization scheme has other interesting applications. \zfr{nonlinear-chain}(a) shows a particular example of a \I{non-linearly} spaced spin array, with all couplings included. Note that even if one were to calculate the appropriate periods $(\B{\qt_z},\B{t_m})$ for engineering the modified chain, it would be difficult to refocus only the nearest-neighbor couplings since the usual grating function ${\cal G}_{ij}$ consists of linearly-spaced peaks. An option would be to use a small value of $N$, leading to broad peaks; however, this has the disadvantage of greater amplitude in the sidelobes (see \zfr{filter-charac}) and poor decoupling of long-range interactions (especially in more complex networks). Sinc-apodization provides a useful workaround by allowing the use of large $N$ (and consequently low side-band power), and broader grating maxima that might refocus the NN couplings. \zfr{nonlinear-chain}(b) demonstrates that reasonably effective Hamiltonian engineering, and high transport fidelities can be achieved even for such non-linear spin arrays.

\section{Off resonance DQ excitation to construct $U_z$}
In the main text, we assumed that the free evolution
propagators $U_z$ were constructed by employing a homonuclear decoupling sequence
(for eg. WAHUHA ~\cite{Waugh68}) during the $\qt_z$ intervals to refocus the internal Hamiltonian couplings, while leaving the action of the gradient. 
An alternative method is to use the DQ excitation sequence even during these periods, but shifting the offset frequency
$\xo_0$ far off-resonance. This is possible since the
recoupling filter is periodic with a period $2m\pi$ thus we can
shift $\xo_0$ while  retaining {identical} filter characteristics.
\zfr{offset} shows that  an offset  a few times larger than the coupling is sufficient to
reach almost free evolution. Still, employing a decoupling sequence
requires almost the same control requirements, and has the added advantage of
enhancing the net decoherence time of the system, allowing for transport in
longer engineered spin chains.
\begin{figure}[h]
 \centering
\includegraphics[width=0.5\textwidth]{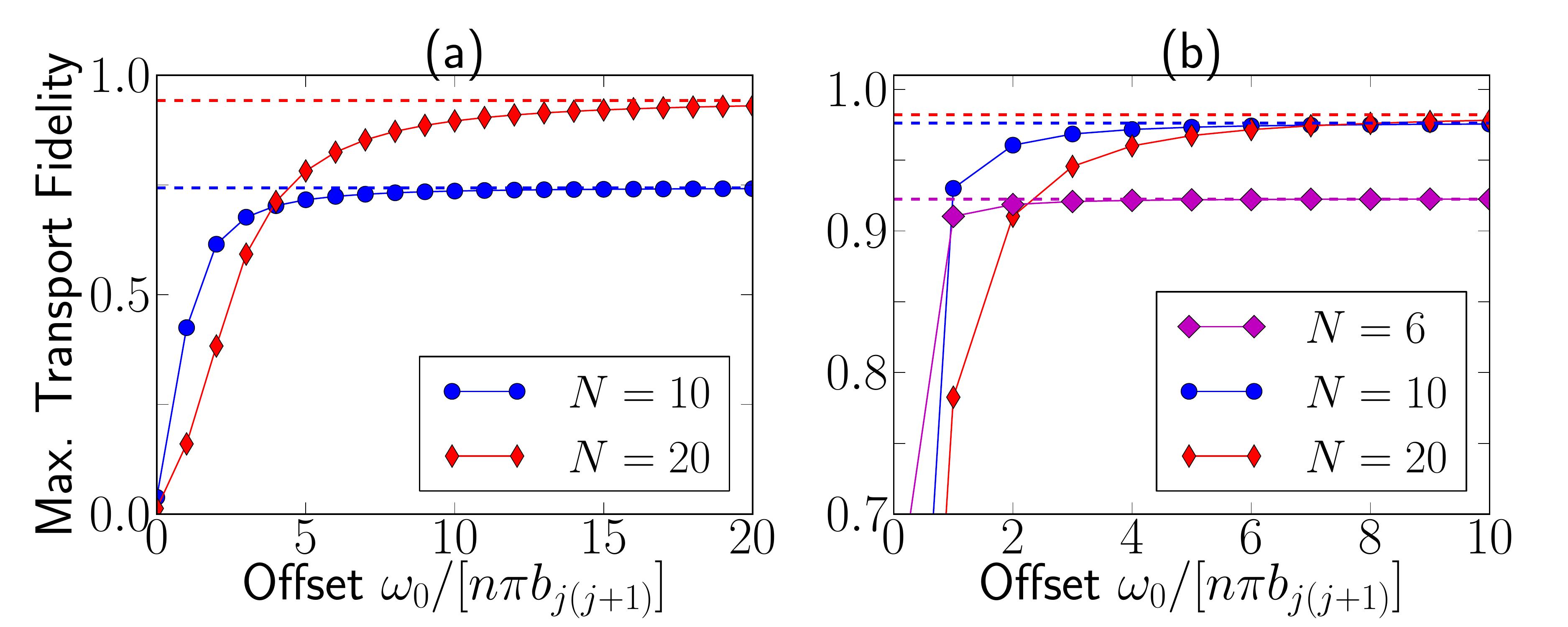}
\caption{A possible implementation of the $U_z$ blocks is by shifting the offset
frequency of the DQ excitation $\xo_0$ to $mn\pi b_{j(j+1)}$. Secular truncation
ensures
that even an offset by a few filter periods is sufficient to obtain fidelities
(solid lines) comparable to when the $U_z$ blocks are created by dipolar
decoupling (dashed lines). We considered in (a) the network in Fig. 1 of the
main text and in (b) a 9-spin chain
with all couplings. 
}
\zfl{offset}
\end{figure}

\begin{figure}[th]
 \centering
\includegraphics[width=0.5\textwidth]{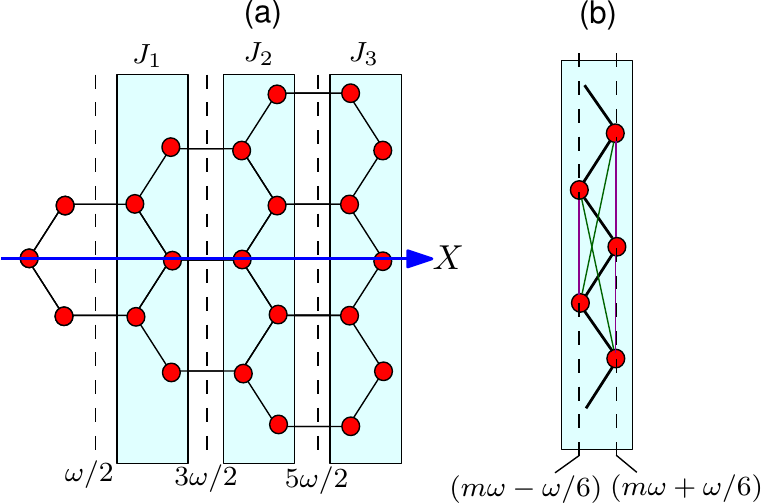}
\caption{Extension of Hamiltonian engineering to other spin lattices. (a) Left panel shows a honeycomb lattice, and the blue line along the X axis denotes the gradient direction. The dashed lines denote the magnetic field strengths along the gradient. It is possible to extract from the the \I{fully-coupled} honeycomb lattice, arrays of nearest-neighbor spin chains (blue boxes). Moreover, the relative nearest-neighbor coupling strength of the different chains $J_j$ can be engineered by filtered engineering. (b) The right panel describes the principle of operation. The nearest-neighbor spins of the chain are at field positions $m\xo \pm \xo/6$, where $m$ is an integer. When $\qt=\pi/\xo$ in the Hamiltonian engineering sequence, NN couplings (dark black) satisfy the Bragg maxima condition and are refocussed. The NNN couplings (magenta) on the other hand pick up a phase $2m\pi \pm 2\pi/3$, and are decoupled when $N>6$ cycles. The decoupling efficiency is limited by the NNNN couplings (green), which are 18.52 times weaker than the NN couplings.}
\zfl{hex-lattice}
\end{figure}
\section{Extension to more complex lattices}
The techniques developed in this paper are not limited to spin chains, but can be directly extended to engineering more complex spin lattices. \zfr{hex-lattice}(a) shows one such example of a honeycomb lattice (naturally occurring for example in graphene). Assuming a \I{fully-coupled} lattice where spins are coupled by the dipolar interaction, we demonstrate below how one could ``extract'' from this lattice, parallel nearest-neighbor-only chains of tunable NN coupling strength. A more detailed extension of the filtered Hamiltonian engineering method to 2D and 3D lattices will be presented elsewhere~\cite{unpub}.

Consider the lattice of  \zfr{hex-lattice}(a), and assume a linear gradient of strength $\xo$ is placed along the X axis (blue line). \zfr{hex-lattice}(b) details the principle of operation. When the Bragg condition is satisfied with $\qt=\pi/\xo$, the NN couplings (black) are perfectly refocussed. Following \zfr{filter-charac}, to a good approximation, the NNN couplings are decoupled if we employ greater than 6 cycles. Of course, the NNNN couplings remain refocussed. However for the honeycomb lattice they are 18.52 times weaker than the NN couplings, and this factor sets the decoupling efficiency of the method. Note that different parallel chains can be engineered to have different relative NN coupling strengths $J_j$, by a suitable choice of the weighting function $F_{ij}$.

\end{document}